\documentclass[a4paper,num-refs]{oup-contemporary}

\journal{gigascience}

\usepackage{graphicx}
\usepackage{siunitx}

\usepackage[normalem]{ulem}

\title{Key challenges facing data-driven multicellular systems biology}

\author[1,\authfn{1}]{Paul Macklin}

\affil[1]{Department of Intelligent Systems Engineering, Indiana University, Bloomington, IN USA}

\authnote{\authfn{1}macklinp@iu.edu}

\papercat{Review} 

\runningauthor{P. Macklin}

\jvolume{00}
\jnumber{0}

\newcommand{\red}[1]{\textcolor{red}{#1}}
\renewcommand{\red}[1]{#1}

\newcommand{\comment}[1]{}

\newcommand{\done}{\red{\textbf{[done]}}}
\renewcommand{\done}{}

\newcommand{\strike}[1]{\red{\sout{#1}}}
\renewcommand{\strike}[1]{}

\jyear{2019}

\begin{document}

\begin{frontmatter}
\maketitle
\begin{abstract}
Increasingly sophisticated experiments, coupled with large-scale computational models, have the potential to sys\-tem\-ati\-cally test biological hypotheses to drive our understanding of multicellular systems. 
In this short review, we explore key challenges that must be overcome to 
achieve robust, repeatable data-driven multicellular systems biology. 
If these challenges can be solved, we can grow beyond the current state of 
isolated tools and datasets to a community-driven ecosystem of 
interoperable data, software utilities, and computational modeling platforms. 
Progress is within our grasp, but it will take community (and financial) commitment. 
\end{abstract}

\end{frontmatter}


\section{Background}
In the past decade, we have seen tremendous advances in measuring, annotating, analyzing, understanding, 
and even manipulating the systems biology of single cells. Not only can we perform single-cell multi-omics  measurements in high throughput 
(e.g., \citep{multiplex-IHC,multiomics-ecoli,multiomics}), but we can manipulate single cells (e.g., by CRISPR systems \citep{CRISPR}), and we can track single-cell 
histories through novel techniques like DNA barcoding \citep{DNA-barcoding}. 

As these techniques 
mature, new questions arise: How do single-cell characteristics affect multicellular systems?   
How do cells communicate and coordinate? How do systems of mixed cell types create 
specific spatiotemporal and functional patterns in tissues? How do multicellular organisms cope with single-cell mutations and other errors? Conversely, given a set of functional design 
goals, how do we manipulate single-cell behaviors to achieve our design objectives? 
Questions like these are at the heart of \emph{\textbf{multicellular systems biology}}. 
As we move from understanding to designing multicellular behavior, we arrive at 
\emph{\textbf{multicellular systems engineering}}. 

High-throughput multiplex experiments are poised to create 
incredibly high-resolution datasets describing the 
molecular and behavioral state of many cells in three-dimensional tissue 
systems. Computational modeling---including dynamical simulation models 
and machine learning approaches---can help make sense of these data. 

Modelers ``translate'' a biologist’s current set of hypotheses 
into simulation rules, then simulate the system forward in time.  
They compare these results to experimental data to evaluate the hypotheses, 
and refine them  until simulations match experiments \citep{PhysiCell_EMEWS,EMEWS2}. 
Computational models allow us to ask ``what if'' questions 
\cite{macklin17_cell_systems}.  
\emph{What if} we added a new cell type to the mix? \emph{What if} we 
spliced in a new signaling pathway? How would our system change? 

Machine learning and bioinformatics complement the dynamical modeling approach: 
analyses of large datasets---especially when annotated with expert-selected 
biological and clinical features---can be mined to discover new relationships between 
single-cell states and behaviors, multicellular organization, and emergent 
function. This, in turn, can drive new hypotheses in simulation models. 
Moreover, machine learning can provide novel analyses of simulation data, 
increasing what we learn from the efforts. 

\begin{figure*}
\includegraphics[width=\textwidth]{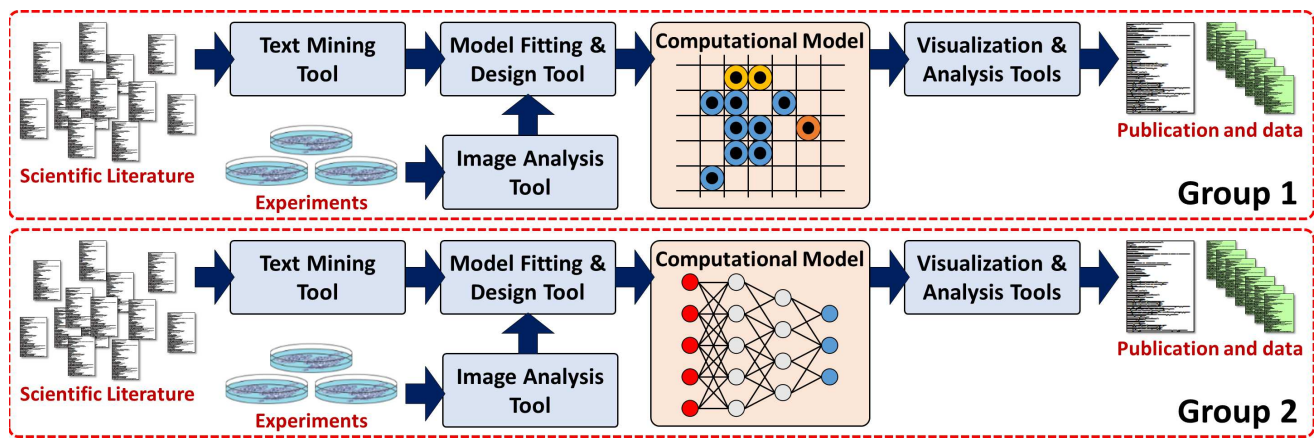}
\caption{Currently, data-driven workflows are largely parallel, with custom-made, incompatible data and tools.} 
\label{fig1}
\end{figure*}

Examples of these approaches appear largely as isolated efforts. 
Most groups seek out their own data sources (previously published data and 
tailored experiments), 
build their own models, and perform their own analyses. Much of this work uses in-house tools created to work on datasets with \emph{ad hoc}, non-interoperable 
data elements. See Figure \ref{fig1}. Thus, any one group's work is by and large incompatible with any 
other group's, hindering or altogether preventing replication studies and modular reuse of 
valuable data and software\red{.}\strike{tools.}

\begin{figure}
\includegraphics[width=0.48\textwidth]{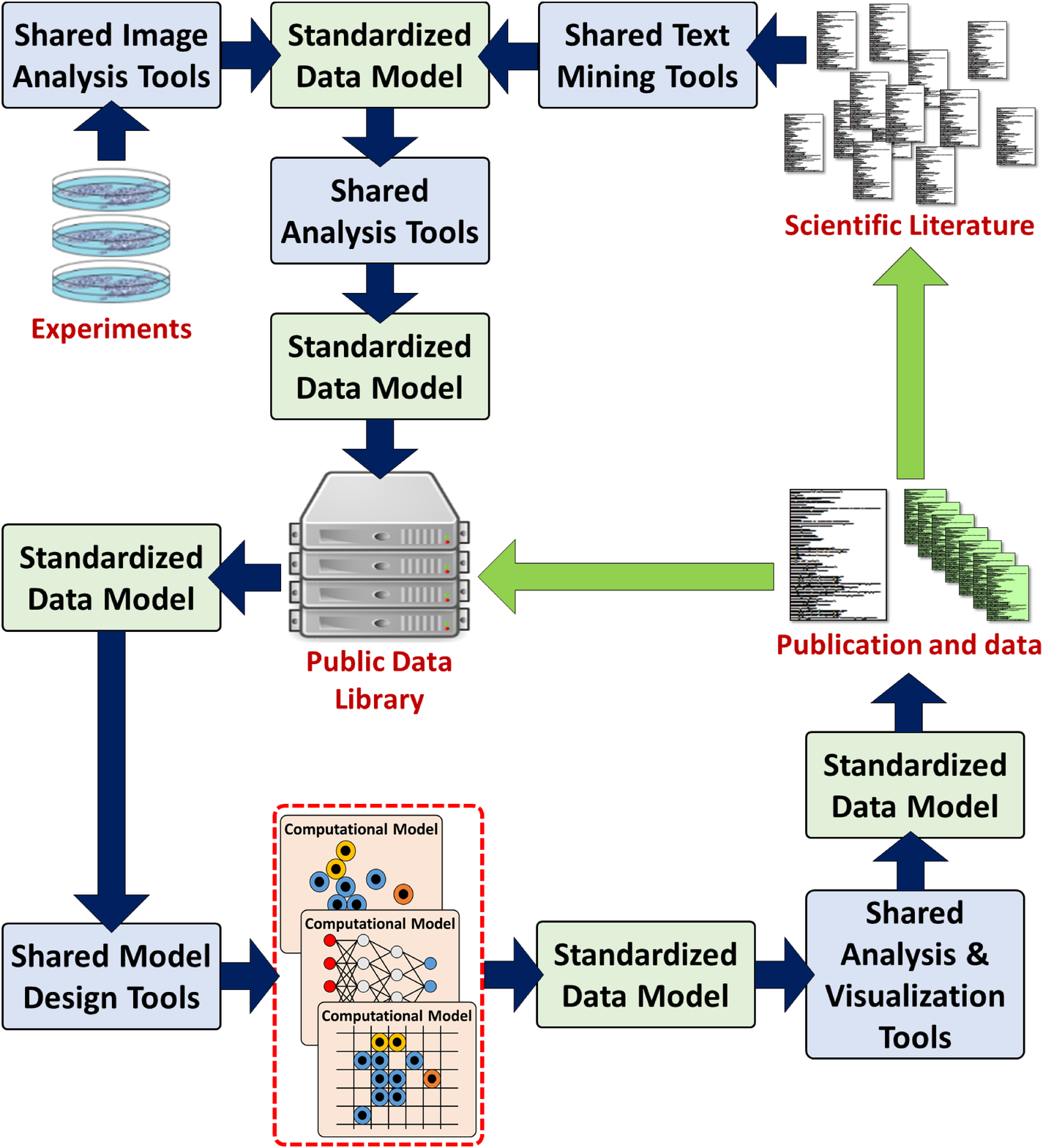}
\caption{If the community can overcome key challenges, an ecosystem of interoperable computational modeling, analysis, configuration, visualization, and other tools could work on community-curated data 
and aggregate insights from many sources.} 
\label{fig2}
\end{figure}

It doesn't have to be this way. If we could solve key 
challenges, we could move beyond single-lab efforts to a community built around 
compatible data and software. Multiple experimental labs could pool their efforts 
to characterize common experimental model systems, and record their data 
in centralized repositories. With a shared ``data language,'' labs could cooperatively 
build better simulation, analysis, and visualization tools. 
Multiple computational labs could build models off of these shared data and tools, find new biological insights, and feed them back into the community. See Figure \ref{fig2}.

In this review, we will explore \red{some} key challenges that \red{we need to} \strike{must be} overcome 
before we can \red{reach the full potential of } \strike{create} an ecosystem of 
interoperable data and tools for multicellular systems biology. 

\red{While the challenges are not presented in any ranked order of importance 
or priority, they progress from the concrete challenges of standardized data 
representation and knowledge capture to  community resources we could build with 
standardized data.  We do not need to address these challenges sequentially. 
One of the 
great strengths of open research communities is that progress can occur by many 
groups in parallel, each contributing according to their individual skills, resources, and interests. 
}


\section{Key Challenges}

\subsection{1. Shared multicellular data standards} 
\label{challenge1}
Data arising from high-throughput experiments 
need to be machine readable and 
stored in interoperable formats with biologically meaningful data elements. 
We need to move beyond shared drives of raw images 
and spreadsheets, to extracted biological 
data elements that are useful for building models and machine learning. 
We need to store not only averaged cell data, but also 
single-cell states for many cells at multiple time points. 
Measurements lose meaning without context: 
data must be stored 
with metadata including detailed cell line and (molecular) growth media 
details, biophysical culture conditions, who performed the measurements, what instruments were used, and what software tools were used for  analysis. 

\subsubsection{Current progress} Great strides have been 
made towards this challenge. The Open Microscopy Environment (OME) 
has emerged as a biological image standard with extensive metadata \citep{OMERO}, 
which has helped to make scientific instruments more interoperable. 
The ISA-Tab format \citep{isatab} functions as a rich online file system: 
provenance and other metadata are bundled with raw data of any file type, 
allowing the contents to be indexed and searched without  detailed 
knowledge of the data formatting. 
This has facilitated the creation of large databases of very 
heterogeneous data (such as GigaDB \citep{gigadb}), and it enables 
simple data exchange due to its support for many data types. 

While these formats facilitate file-level interoperability, they do not 
encode extracted biological data elements. Protocols.IO 
was developed to share detailed experimental 
protocols \citep{ProtocolsIO}, which can be cited in journal 
publications to help improve repeatability and reproducibility. 
However, the protocols are human-readable checklists; 
they do not use a machine-readable controlled vocabulary of
growth factors and other culture conditions. 

Ontologies such as the Medical Subject Headings (MeSH) \citep{mesh1,mesh2} and 
the Cell Behavior Ontology (CBO) can annotate many biological concepts \citep{CBO}, 
but they serve as controlled vocabularies rather than 
standardized data formats.

The Systems Biology Markup Language (SBML) is a well-established 
standard for single-cell systems biology \citep{SBML}, and efforts such as SBML-Dynamic 
are working to extend SBML to multicellular models. Domain experts in computational biology, 
experimental biology, and data science worked together to draft MultiCellDS, a standard 
for multicellular data \citep{MultiCellDS}. MultiCellDS has a 
highly extensible representation of single-cell phenotype built from a 
variety of ontologies such as MeSH and CBO, which can be used 
to represent highly multiplex data (e.g., \citep{multiplex-IHC}) for 
many cells, along with metadata and microenvironmental context. 
\red{The European Union-funded MULTIMOT project has been developing 
a community-driven standard for cell motility measurements 
(MIACME: Minimum Information about Cell Migration Experiments), with a 
corresponding software ecosystem \cite{MULTIMOT} that can interface with 
data in ISA-Tab and OME formats.} 

\subsubsection{Future} None of these efforts has completely addressed this challenge. 
Ultimately, we should combine and extend them into a unified data format. 
ISA-Tab could bundle image data (using OME) and 
extracted biological features (e.g., with MultiCellDS \red{and MULTIMOT}), while storing  
experimental protocol details with a controlled vocabulary growing out of 
Protocols.IO \cite{ProtocolsIO}. 

\red{We must ensure that metadata not only annotate experimental protocols, 
but also data extraction protocols: What algorithms were used to extract the 
biological data elements, and where is the source code permanently archived? 
Some popular data science software (e.g., Docker and Jupyter notebooks) allow 
users to export their computational pipelines facilitate this reproducibility. 
Lastly, note that extracted biological data elements cannot \emph{replace} 
raw data: end users must be free to reproduce (and improve!) the extraction 
of data elements, which requires access to the original data.}

\subsection{2. Shared multicellular observational representations}
\label{challenge2}
Beyond quantitative measurements like cell division rates, 
we need a machine-readable encoding of qualitative observations 
and insights derived from raw biological data: 
when cells are in condition \emph{X}, they do 
\emph{Y}. When cells of type \emph{X} and \emph{Y} interact by contact, 
they tend to do \emph{Z}. When cell line \emph{X} looks like \emph{Y} in an 
experiment, the cell culture medium lacked factor \emph{Z}. 

Labs and clinics are replete with such examples of 
hard-won knowledge, but until we can systematically record them, 
these insights will remain siloed, isolated, and destined to be relearned, 
lab by lab. 
If we could consistently record qualitative observations, we could progress from 
single-cell measurements to multicellular systems understanding, including annotation of critical cell-cell interactions. 

Until we can specify ``correct'' model behavior with machine-readable 
annotations, our simulation studies will be rate-limited to how quickly humans 
can view simulations and assess them as more or less ``realistic.'' 
How do we say, in a generalized way, that a simulated tumor 
stays compact or becomes invasive? How do we 
know if a simulated developmental process has the ``right'' amount of branching? 
What does it mean for simulated image \emph{X} to ``look like'' experimental image \emph{Y}, given that both the simulation and the experiment are single instances of stochastic processes? If we cannot record the qualitative behavior of 
simulations and experiments, we cannot automate processes to compare them. 

\subsubsection{Current progress} Progress on this challenge has been limited. 
The CBO \citep{CBO} has developed a good starting 
vocabulary for observed cell behaviors. Extensions of SBML \citep{SBML} 
could also potentially 
represent some of these multicellular and multiscale observations. Tailored 
image processing has been applied to individual 
investigations to extract (generally quantitative) representations, although 
to date we have seen few (if any) qualitative descriptors generated by systematic 
image analysis.

\red{There has been greater progress in presenting phylogenetic relationships in 
multicellular populations with automatically extracted phylogenetic trees 
and other data visualizations, such as Muller plots (e..g, \cite{SPADE,hicks_tree,evofreq}). 
These techniques examine large multiomics datasets (e.g., scRNA-seq data \cite{scRNA_seq}) 
to fit and represent lineage relationships between cell types (or classes) 
with directed graph data structures.}

\subsubsection{Future} This area \strike{seems} \red{is} ripe for machine learning: given a set of qualitative descriptors 
like ``compact'' versus ``invasive,'' ``mixed'' versus ``separated,'' ``growing'' versus 
``shrinking'' or ``steady,'' a neural network could be trained to human 
classifications of experimental and simulation data. High-throughput 
multicellular simulators (e.g., \citep{PhysiCell_EMEWS})  
could create large sets of training data in standardized formats 
with clear ground truths. Machine vision 
could also be be used to analyze time series of multicellular 
data. These annotations could give rise to metrics that help us 
systematically compare the behavior of one simulation with another, 
or to determine which simulation (in a set of hundreds or thousands of simulations) 
behaves most like an experiment. 

\red{Graph structures could also be applied to represent and visualize 
cell-cell interactions in multicellular populations \cite{MultiCellDS}, 
similarly to phylogenetic trees (e.g., \cite{hicks_tree,evofreq}), chemical reaction networks 
(e.g., \cite{graph_chem_network,graph_chem_network1}), 
gene network diagrams (e.g., \cite{gene_network_graph}), 
and emerging data formats for agent-based model rules (e.g., as in Morpheus \cite{Morpheus}).}

\subsection{3. Standards support in computational tools}
\label{challenge3}
For data standards to be truly useful, they must be broadly supported by a 
variety of interoperable tools. 

\subsubsection{Current progress} Single-cell systems biology has already shown the 
enabling role of stable data standards \citep{opensource-review}: once SBML 
crystallized as a stable data language, a rich and growing 
ecosystem of data-compatible simulation and analysis software emerged. 
Multicellular systems biology has not yet reached this point: 
most computational models have custom configuration and output formats, 
sometimes with customized extensions of SBML to represent single-cell systems 
biology \citep{MultiCellDS}. 


\subsubsection{Future} If a multicellular data standard emerges, key open source projects \citep{opensource-review} can implement read and write support in their software, either ``natively'' (i.e., 
at run-time), or as data converters. Hackathons or similar hosted workshops could 
facilitate  this work. Ontologists need to provide user-friendly data bindings to simplify these development 
efforts. If standards are to be supported more broadly than just major 
open source packages,  we must remember that most scientific software 
is created with little formal 
software engineering training;  the data bindings must be well-documented, 
have simple syntax, and require minimal   installation effort.

\subsection{4. Shared tools to configure models and explore data}
\label{challenge4}
It is not enough to simply read and write data into individual tools. 
We must reverse the current ``lock in'' effect: because 
multicellular modeling software is difficult to learn, users (and often entire labs) 
focus their training  on a single modeling approach. Because of this, replication studies 
are rare, even when a study's source code and data are openly available. 

To solve this, we need user-friendly tools to import and set biological and 
biophysical parameters, design the virtual geometry, and write standardized 
configuration files that initialize many modeling frameworks. Users could 
run models in multiple software packages, replicate the work of others, and 
avoid software-specific artifacts that can bias their conclusions.

Shared software to read, analyze, compare, and visualize outputs from 
multiple modeling packages could reduce the learning curve for new software. 
If the shared data exploration and analysis tools were written to work on a 
common format that includes segmented experimental data, they could 
also be used to explore experimental data, make and annotate new observations, 
and motivate new model hypotheses \cite{assaf_resuse_images}. 

\subsubsection{Current progress} 
Without a common format for multicellular simulation data, there has been little
opportunity to develop shared tools for configuring, running, and visualizing  multicellular simulations. 
Some individual simulation packages such as Morpheus \citep{Morpheus} and CompuCell3D \citep{CC3D} have 
user-friendly graphical model editors, but they are currently limited to their 
individual user communities and not compatible with other simulation packages 
\citep{opensource-review}. 
Commercially-backed open source software such as Kitware's 
ParaView \citep{paraview} 
is commonly used to visualize multicellular simulation data, but only by writing 
customized, simulation-tailored data importers. ParaView is generally 
not used to visualize biological data. 

\red{Cloud-hosted tools have provided a means to share 
sophisticated tools with broad, multidisciplinary audiences 
without need for downloading and compiling the tools. 
For example, the National Cancer Institute (NCI) has 
introduced NCI cloud resources as part of the NCI Cancer 
Research Data Commons \cite{NCI_cloud_resources}. 
Sophisticated simulation models can also be 
shared as web applications: the PhysiCell development 
team recently created xml2jupyter \cite{xml2jupyter} to 
automatically create Jupyter-based graphical user interfaces (GUIs) 
for PhysiCell-based multicellular simulations, which 
can then be cloud-hosted on platforms like nanoHUB \cite{nanohub}. } 

\red{Other model and data sharing paradigms that emerged to 
address related issues in reproducibility may also encourage reuse, such as 
bundling data and software with Binder \cite{binder} or 
\emph{GigaScience}'s recent partnership with CodeOcean 
to pair papers with cloud-hosted executable platforms 
\cite{gigascience_code_ocean}. However, these typically are 
single-purpose workflows (specialized to a specific data analysis for a single 
paper) that are not designed for modular reuse 
in new research workflows. They tend to lack standardized data formats 
(see Challenges 1-2) to facilitate connection 
with other tools, and latency issues will challenge their use 
use in high-throughput workflows. Moreover, note 
that while cloud-hosted executable codes increase accessibility and 
availability, they must not substitute for (or circumvent) 
sharing source code for full reproducibility.}

\subsubsection{Future} It will be difficult to make progress on this 
challenge without stable standards for multicellular input and output data. 
However, progress could be made using current draft standards, such as 
MultiCellDS \citep{MultiCellDS}. ParaView could use customized plugins to 
support emerging standards for multicellular data. If 
 projects like Morpheus implemented standards, 
their graphical model editors could become valuable community resources. 

Hackathons can help to rapidly prototype new tools (particularly 
if they are paired with benchmark datasets), but they 
must aim to create well-documented, engineered software that are 
maintained in the long term. 
We may need new funding paradigms to support small open source teams. 
\red{The form of these funding paradigms is not fully clear. Hackathons and 
similar forms of focused, small-team collaboration could possibly be sponsored  
through existing federal and philanthropic mechanisms for meetings and travel 
grants. Crowdsourcing could potentially fund some focused community tool 
development and maintenance. There is also room for creativity among 
funding organizations for smaller grants with faster review cycles for 
community tool building efforts. }

\red{Lastly, shared code platforms such as the NCI Data Commons could provide 
an environment  to connect data and tools in online, easy-to-use workflows
that encourage scientists to ``mix and match'' data software components 
into unique research. However, it will be important to avoid ``lock-in'' effects 
that prevent moving data and tools from one platform to another. 
Moreover, as workflows come to incorporate more web services (in differing platforms), 
they could become vulnerable to technical failures, business failures, or 
malicious attacks. Open source software has largely solved these issues 
by mirroring software repositories. Web services may need similar mirroring, 
and open science norms will need to encourage source code sharing and data/tool 
portability for web platforms just as they have for offline code. 
} 

\subsection{5. High-quality, multiscale benchmarking datasets}
\label{challenge5}
Once we have standardized data formats and an ecosystem of compatible 
software to support them, we need high-quality datasets to drive the 
development of computational models. The ideal 
datasets would sufficiently resolve single-cell morphologies and 
multi-omic states in 3-D tissues, along with microenvironmental 
context (e.g., spatial distribution of oxygen).

To capture the behavioral states of cells, we need standard 
immunohistochemical 
panels that capture multiple dimensions of cell phenotype: cycle status, 
metabolism, death, motility (including markers for the leading edge), 
adhesiveness, cell mechanics, polarization, and more. We will need to
capture these details simultaneously in many cells  
at multiple time points, using massively multiplexed technologies. 

These datasets would be used to formulate model hypotheses and assumptions 
(through data exploration using standardized tools), to train models, 
and to evaluate them. Moreover, as the community develops new 
computational models,  
they could be evaluated against benchmark datasets. 
Benchmark datasets are domain-specific: 
separate datasets are needed for developmental biology, 
avascular and vascular tumor growth, autoimmune diseases, and other problems. 
It is important that these datasets are easily accessible with open data licenses 
to promote the broadest use possible. Adhering to 
FAIR (Findability, Accessibility, Interoperability, and Reusability) data 
principles would be ideal \citep{FAIR}. 

\subsubsection{Current progress} 
Cancer biology has made perhaps the greatest progress on this challenge,  
where the NIH-funded 
Cancer Genome Atlas 
hosts many genomic, microscopy, and other large datasets \citep{TCGA}.
Typically, these 
consist of many samples at a single  time, rather than 
time course data. Highly multiplex multicellular data are generally not 
available. 
DREAM challenges have 
assembled high-quality datasets to drive model development 
(through competitions) \citep{dream-challenges}, but these have not typically satisfied 
the multiplex, time series ideals outlined above. Private foundations 
are using cutting-edge microscopy to create high-quality online datasets 
(e.g., the Allen Cell Explorer Project \citep{allen-cell-explorer}). 

The technology for highly-multiplexed measurements is steadily improving: 
CyTOF-based immunohistochemistry (e.g., as in \citet{multiplex-IHC}) can 
stain for panels of 30-50 immunomarkers on single slides at 1-2 
$\mu\textrm{m}$ resolution or better. There are  no standardized
panels to capture the gamut of phenotypic behaviors outlined above. Social 
media discussions 
(e.g., \citep{twitter-motility}) %
have helped to drive community dialog on difficult phenotypic parameters, 
but no clear consensus has emerged for a ``gold standard'' panel of 
immunostains. 

\subsubsection{Future}
Workshops of leading biologists should assemble the ``dream panel'' 
of molecular markers. Consortia of technologists will need to reliably implement these multi-parameter panels in experimental workflows \cite{multiplex-IHC}. 
Workshops of bioinformaticians, data scientists, and 
modelers will be needed to ``transform'' these raw data into standardized datasets for use in models. All this will require federal or philanthropic funding, and contributions by multiple labs. 
Social media has great potential for public brainstorming, 
disseminating resources, and recruiting new contributors. Hackathons could 
help drive the ``translation'' of raw image data into standardized datasets, while developing tools that  automate the process. 

\subsection{6. Community-curated public data libraries}
\label{challenge6}
We need ``public data libraries'' to store and share high-quality, standardized data \cite{public_archives,assaf_resuse_images}. 
Data should not be static: the community should continually update data to reflect 
scientific advances, with community curation to ensure data quality. 
Public libraries must not only store 
raw image data and extracted biological parameters, but 
also qualitative observations and human insights. 
The public libraries should host data 
at multiple stages of publication: preliminary data (which may or may not be 
permanently archived), datasets under construction (i.e., the experiments 
are ongoing), data associated with a preprint or a paper in review, 
and data associated with a published work. 
Public data libraries should \red{enable if not} encourage versioned post-publication 
refinement, \red{particularly for datasets arising from secondary analysis or 
curation of heterogeneously sourced primary raw data, such as digital cell lines \cite{MultiCellDS}}. 
Lastly, public data libraries need to be truly public by using 
licenses (e.g., Creative Commons CC0 or CC-BY) that encourage 
new derivative works, as well as aggregation into larger datasets. 

\subsubsection{Current progress} 
Numerous data portals exist, and more are emerging. Many are purpose-built 
for specific communities, such as the 
Cancer Genome Atlas \citep{TCGA}. 
\red{The Image Data Resource \cite{IDR} was recently launched to facilitate 
sharing bioimages using the OME data format \cite{OMERO}, further demonstrating 
how standardized data can facilitate the creation of shared tools and resources.}
Others like GigaDB \citep{gigadb} 
and DRYAD \citep{DRYAD} 
allow users to post 
self-standing datasets with unique DOIs to facilitate data reuse 
and attribution. 
These repositories  are free for access, 
thus increasing the reach and impact of hosted data, but the 
data contributors must pay at the time of data publication. 
The fees often include editorial and technical assistance while 
ensuring long-term data availability. 

Even within single data hosting repositories, individual datasets are largely disconnected 
and mutually non-interoperable beyond ISA-Tab compatibility. 
Thus, individual hosted datasets and studies are generally not 
bridged and recombined. Moreover, the datasets are usually  static after 
publication, rather than actively curated and updated. 
BioNumbers 
has long served as a 
searchable resource of user-contributed biological parameters \citep{bionumbers}, but it lacks a 
unified data model.  The MultiCellDS project proposed 
\emph{digital cell lines}, which aggregate measurements from many sources 
for a single cell type \citep{MultiCellDS}. Digital cell lines were intended to 
be continually updated and curated by the community, so that low-quality measurements 
could be replaced by better measurements as technology advances. 
However, this effort is currently  manual, with no single, easily searchable 
repository for its pilot data. 

An unfortunate consequence of the current data hosting model is that all the 
burden rests on data donors: they generate the data, format it to standards, 
assemble it, document it, upload it, and then pay the hosting and 
scientific publication costs. This is a classic case of the \red{``\emph{tragedy of the 
commons}''}: it is easy to benefit from shared resources, but 
\red{the cost of contribution falls on contributors.} \strike{costly to contribute.} 
Most repositories have fee waivers  for scientists in low-income 
nations, but small and underfunded labs and citizen scientists are 
still at a disadvantage.

\red{Nonprofit organizations like DRYAD have made great strides in 
creating sustainable resources to host data; currently (as of 2019), 
a one-time charge of US \$120 per dataset applies once the data are accepted by 
curators and publicly available \cite{DRYAD}. This is a small fee compared to the 
data generation cost for experimental labs and within the means of 
well-funded labs. In cases where secondary analyses or simulations generate 
new datasets independent of grant funding, there may be greater hardship 
in these costs, particularly when coupled with open access publication fees.} 


\subsubsection{Future}
We need to develop more unified, \strike{financially stable and} scalable 
repositories that can bridge fields and collect our knowledge. 
The repositories \red{should} \strike{need to} be \red{indexed and} community curated \strike{and continually improved, 
rather than static} \red{to encourage continuous refinement where possible.} 
\red{While there has been great progress to create financially sustainable, permanent data hosting, 
there is still room to explore alternative funding for data generated 
independently of specific grant funds. Moreover, these archive-oriented data stores still 
require curation and indexing if they are to grow from data storage to libraries.}


Solutions to this 
challenge may well originate outside the bioinformatics community. 
Library scientists have longstanding domain expertise in collecting and 
curating knowledge across disciplines in unified physical libraries: this expertise 
would undoubtedly benefit any efforts to create public data libraries. The tremendous 
success of Wikipedia \citep{wikipedia} 
in hosting its 
own image and video resources on Wikimedia Commons \cite{wiki-commons}---%
at no cost to contributors---could be a very good model. 
bioRxiv \citep{biorxiv} 
has been similarly successful in hosting preprints at no cost to authors\red{, although 
experimental data hosting costs are far higher than the cost of hosting manuscripts}. 
Both of these have relied upon a combination of public donations, federal support, 
and philanthropy, channeled through appropriate nonprofit structures. 

\red{We note that public data libraries could become victims of their own success: as 
public repositories proliferate, finding information will become increasingly difficult, 
and the community of contributors could become fragmented. This, in turn, will 
make it difficult to recruit data curators to maintain the quality of the resources. 
Thus, the community may need to reach consensus on which libraries serve as the 
standard repositories for which types of data. Moreover, unified search engines 
and indices may be needed to help unify knowledge in existing and new data libraries.}

Lastly, to ensure robustness and sustainability, we need to encourage data 
mirroring with global searchability, and promote a culture that values and properly 
cites all contributions to shared knowledge: data generation, data analysis, 
and data curation. While badges can help \citep{badges1,badges2}, we must ensure that 
data users can easily cite all these contributions in papers, that impact metrics 
reflect the breadth of contributions, and that tenure and other career processes truly 
value all contributions to community knowledge resources. 

\subsection{7. Quality and curation standards}
\label{challenge7}
Community-curated public libraries face new questions: how can we consistently decide 
which data are worth saving? How do we determine if a new measurement is 
better than an old one? How do we monitor quality? Can we automatically trust 
one lab's data contributions based upon prior contributions? And who gets to make 
these decisions? 

\subsubsection{Current progress} Little to none, aside from   
uncertainty quantification.

\subsubsection{Future} This challenge is as much cultural as it 
is technical. We will need to 
hold workshops of leading biologists to identify community values and standards for assessing 
different measurement types.  The community will need to determine 
if ``gold standards'' can be devised for comparing measurements. 

\subsection{8. Linking data to models}%
\label{challenge8}
We need to connect data to computational models. Data modelers should 
help design experiments, to determine 
what variables are needed to build useful models. 
We need to determine how to ``map'' biological measurements to model parameters.

\subsubsection{Current progress} This challenge is currently being 
addressed on a study-by-study basis. Individual 
teams design  experiments, devise their own model calibration methods, formulate 
 model evaluation metrics, and create their own tools to analyze 
and compare experimental and simulation data. 

\subsubsection{Future} 
This challenge is both technical 
and cultural. Mathematicians, biologists, data scientists, and others will 
need to work together to determine what it means for an inherently stochastic 
simulation model to match to match an experiment. Any progress in 
creating standardized data elements 
and annotating multicellular systems behaviors 
will surely help in creating metrics to compare experimental and computational 
models. Once standardized biological parameters are extracted to create 
benchmark datasets, 
machine learning could help drive more systematic mappings from extracted biological parameters to 
computational model inputs. 
 
\section{Conclusions}
The time is ripe for data-driven multicellular systems biology and engineering. 
Technological advances are making it possible to create high-resolution, 
highly multiplex multicellular datasets. Computational modeling platforms---including 
simulation and machine learning approaches---have advanced considerably, and they 
are increasingly available as open source \cite{opensource-review,multicell-sysbio-progress}. 
Supercomputing 
resources are amplifying the power of these computational models \cite{PhysiCell_EMEWS,EMEWS2}, 
while cloud resources are making them accessible to all \citep{nanohub,xml2jupyter}. 

If we can solve these key challenges, we will connect big multicellular datasets 
with computational technologies to accelerate our understanding of biological systems. 
\red{Steady, incremental progress towards any of the challenges benefit the community as 
we move towards this broader vision.} 


Some of the challenges are largely technical, such as creating data standards. Others 
are more cultural, such as shaping community values for data curation. All 
of the challenges share a need for community investment: developing 
and sharing compatible tools and data, hosting data, curating public data libraries, 
and ultimately funding these worthwhile efforts. Many groups are already 
contributing pieces of this puzzle, often with little financial support. 
In the future, we must reduce the individual burden in creating community goods.
We may need newer, more rapid funding paradigms to help support and harden new  
software tools, scaling from small but simple proposals to the current large software 
grant mechanisms (which tend to have low funding rates). We may need to fund software labs 
rather than software projects, to encourage rapid response to emerging community needs. 

We are on the cusp of accelerated, data-driven biological discovery of how cells work together, 
how they build things, and how this breaks to cause disease. If you are working towards 
solving any of these challenges (or if you have new ones to pose!), please consider sharing your 
advances here.

\section{Declarations}

\subsection{List of abbreviations}

\begin{tabular}{l l}
\textbf{CBO} & Cell Behavior Ontology \\ 
\textbf{CC0} & Creative Commons public domain license\\
\textbf{CC-BY} & Creative Commons attribution license\\
\textbf{CyTOF} & Cytometry by Time Of Flight \\
\textbf{DOI} & Digital Object Identifier \\ 
\textbf{DREAM} & Dialogue for Reverse Engineering Assessments \\
&and Methods \\
\textbf{FAIR} & Findability, Accessibility, Interoperability, \\
&and Reusability\\ 
\textbf{ISA-Tab} & Investigation-Study-Assay tabular format\\ 
\textbf{MeSH} & Medical Subject Headings \\ 
\textbf{NIH} & (U.S.) National Institutes of Health \\
\textbf{OME} & Open Microscopy Environment \\
\textbf{SBML} &  Systems Biology Markup Language \\
\end{tabular}

%

\subsection{Competing Interests \done}
The author declares that he has no competing interests. 

\subsection{Funding \done}
PM's work to develop computational tools and data standards for multicellular systems 
biology was funded by the Breast Cancer Research Foundation 
(PIs Agus, Ewald, Gilkes and Macklin) 
and the Jayne Koskinas Ted Giovanis Foundation for Health and Policy (PIs Ewald, Gilkes\red{, and 
Macklin}), the National Science Foundation (PI Fox, 1720625), and the National 
Cancer Institute (\red{PIs Finley, Macklin, and Mumenthaler, U01-CA232137-01; }
PIs Agus, Atala, and Soker, 1R01CA180149; PI Hillis, 5U54CA143907). 
\subsection{Author's Contributions \done}
PM conceptualized and wrote the manuscript. 

\section{Acknowledgements \done}
PM thanks Nicole Nogoy and the staff of \emph{GigaScience} for the opportunity to 
write this review, and for editorial support. \red{PM thanks the reviewers 
and preprint readers for valuable comments and feedback.} 

\section{Author's information \done}
PM has worked for over ten years in computational multicellular 
systems biology, with a focus on cancer biology and tissue engineering. He has 
written several open source tools for the field, including 
BioFVM (a multi-substrate diffusion solver for biochemical cell-cell communication) 
\cite{biofvm}, PhysiCell 
(a 3-D agent-based modeling toolkit) 
\cite{physicell}, 
and 
MultiCellDS 
(a draft multicellular data standard) \cite{MultiCellDS}. 
He is an Associate Professor of Intelligent Systems Engineering at Indiana University.

\end{document}